# Concise overview of methods to enhance the thermoelectric efficiency of SnTe

Diptasikha Das[1] and Kartick Malik*[2]

[1]Department of Physics, Adamas University, Kolkata, West Bengal, India
[2]Department of Physics, Vidyasagar Metropolitan College, Kolkata, West Bengal
*Corresponding Author Email: kartick.phy09@gmail.com

**Abstract.** SnTe is a potential thermoelectric material (TE) in the mid-temperature range. Detailed techniques to enhance the figure of merit (ZT) by increasing Power Factor (PF), and reducing thermal conductivity ($\kappa$), of SnTe-based TE materials are discussed. The key factors governing the ZT of a TE material are discussed to facilitate the optimization of the efficiency. Various techniques to synthesis bulk and nano-structured SnTe, are presented. Efforts are made to reveal the optimization techniques for ZT of SnTe-based materials through band structure engineering and reduction in thermal conductivity. Nano-structuring is one of the important approaches to decouple the interrelated material properties and reduce $\kappa$. Band structure engineering is employed to enhance the PF.

## 1    Introduction

Energy crisis and environmental pollution are the prime challenges of the 21st-century civilization. The basic form of energy used in industry and society is electrical energy. To date, the main source of electrical energy is fossil fuels. It is crucial to note that 60% of the heat energies, wasted during the conversion to electricity using fossil fuel, and the remaining 40% is employed to generate electricity. However, an affirmative action may be taken to recover the waste heat by employing Thermoelectric (TE) technology. TE device is a solid-state energy converter for heat-to-electrical energy conversion, and vice versa, without the involvement of any mechanical moving parts [1]. The basic principle of TE devices is based on Seebeck and Peltier effects, in the form of thermoelectric generators (TEG) and thermoelectric coolers (TEC). However, the efficiency of the TE material involved in the device may be estimated by the figure of merit [1], $ZT = \frac{S^2\sigma}{(\kappa_e+\kappa_L)}T$. The term $S^2\sigma$ is known as Power Factor (PF), where S, $\sigma$, $\kappa_e$, and $\kappa_L$ are thermopower, electrical conductivity, electronic thermal conductivity, and lattice part of the thermal conductivity, respectively. These are interrelated materials properties, and the efficiency at the commercial level is limited due to low ZT. Worldwide, there are resurgences to enhance the ZT of TE material through decoupling the materials' properties. There are two ways to enhance the ZT: optimization of PF, and a decrease in $\kappa = \kappa_e + \kappa_L$. The general methods for optimization of PF include band engineering, viz., band convergence, resonant doping, band anisotropy, and band nestification etc. [2-3]. However, modification in electrical transport properties through interface engineering and doping are other methods for optimization of PF [2-3]. The other route to enhance ZT, through tuning $\kappa_L$ includes point defects, nano-structuring, scattering at grain boundary, dislocation, hier-archical microstructures, and lattice softening [2-3]. However, the spectrum of TE materials is categorized in three sections, viz., low temperature (T<300K), mid-temperature (500<T<800K), and high temperature (T>900K) TE material based on the temperature at which highest ZT is obtained. Some of the recent, remarkable, and well-known TE materials are Bi2Te3 [4], Mg3Sb2 [5], SnSe [6], PbTe [7], PbS [8], CoSb3 [9], BiCuSeO [10], SnTe [11], GeTe [12], Cu2Se [13], Half-Heusler [14] and SiGe [15].

Mid-temperature TE materials are drawing attention for recovering, industrial waste heat, and heat of vehicle exhaust [16]. Lead chalcogenide, tin chalcogenide, filled Skutterudite, and Half-Heusler based TE materials have the potential to convert waste heat at mid-temperature [9, 18-20]. High ZT~2.5 at 923 K has been achieved for $Pb_{0.98}Na_{0.02}Te-8\%SrTe$, a mid-temperature TE material, by employing valance band convergence, point defects and nanostructuring methods simultaneously [18]. Zhao *et al.* have been observed that SnSe single crystals possess highest ZT~2.6 at 923 K temperature along b axis in orthorhombic unit cell of SnTe at room temperature [19]. Half-Heusler alloy based mid-temperature TE material, $Sr_{0.09}Ba_{0.11}Yb_{0.05}Co_4Sb_{12}$ and Skutterudite based $(In,Sr,Ba,Yb)_yCo_4Sb_{12}$ have been studied, and high ZT ~1.8 are observed at around 835 K [9, 20, 21]. However, Lead telluride (PbTe) and its alloys have been considered as a primary mid-temperature TE material for industry-based systems. The application of PbTe is limited due to the toxicity of Pb [22]. An alternative lead-free compound, SnTe has been introduced as a potential TE material in the mid temperature range. Environment friendly, SnTe is also crystallize as rocksalt structure and electronic structure of SnTe consists of multiple valance bands (light-hole, L and heavy-hole, Σ bands) alike PbTe [22]. SnTe based TE materials: $Sn_{0.94}Ca_{0.09}Te$, $Sn_{0.95}Ag_{0.05}Te_{0.95}I_{0.05}$, $Sn_{0.94}Mg_{0.09}Te$, $Sn_{0.88}Mn_{0.12}Te$, $Sn_{0.98}Mg_{0.03}In_{0.03}Te$, $Sn_{0.85}Sb_{0.15}Te$, $SnCd_{0.12}Te$ etc. have been studied, and high ZT > 1 is observed at around 850 K [23- 25]. Electronic band gap ($E_g$) i.e., gap between conduction band and L band for of pristine SnTe ($E_g \sim 0.18\ eV$) is very small compared to PbTe ($E_g \sim 0.30\ eV$)). However, gap between two valance bands i.e., L and Σ of SnTe ($\Delta E_{L-\Sigma} \sim 0.35\ eV$) is large with respect to PbTe ($\Delta E_{L-\Sigma} \sim 0.17\ eV$) [22]. Intrinsic SnTe possesses high hole concentration ($10^{20}$ - $10^{21}$ cm$^{-3}$) relative to PbTe due to the inherent Sn vacancies. High hole concentration and dissimilar electronic band gap are the inherent reasons for very low *S* and high σ in SnTe, compared to PbTe, cause very poor *ZT* [26]. However, the PF of SnTe may be enhanced by optimization of carrier concentration *n* and band engineering method via valance band convergence, resonance level engineering, synergistic effect and increasing carrier effective mass. Furthermore, the κ may be reduced to optimize ZT by introducing various types of crystal defects viz., point defect, dislocations, interfaces and precipitates in the bulk matrix [27]. However, crystal defects owing to nanostructure is one of the important methods to reduce κ. Nano-sized SnTe, synthesized by hydrothermal method shows low $κ_L \sim 0.6\ wm^{-1}K^{-1}$ and high ZT~0.49 at 803 K [28]. ZT ~ 1.3 at 873 K has been obtained for nano-composite $SnCd_{0.03}Te$ + 2% CdS/ZnS, synthesized by melting process [26].

In this review article, Synthesis processes, including nano-structure synthesis of SnTe based TE materials are discussed. Further, various techniques such as optimization of *n*, band convergence, resonant level, and synergistic techniques are highlighted to optimized PF of SnTe based TE materials. Point defects, nano-structuring and all-scale hierarchical strategies have been discussed for reducing $\kappa_L$. Synthesis of nano-structure and effect of nano-structuring on transport prosperities of SnTe-based TE materials are pivot of the article. Noteworthy, nano-structuring of TE materials, not only reduce $\kappa_L$ but also modify other TE properties.

## 2. SnTe Synthesis

SnTe based TE materials may be synthesized by melting, hot pressing followed by melting, self-propagating high-temperature synthesis (SHS) and plasma activated sintering technique etc. [29]. Further, synthesis of nanosized SnTe has been reported by employing ball milling, hydrothermal, sol-gel, and chemical synthesis method etc. [25, 26]. However, synthesis of SnTe based materials using melting processes is common and mostly cited; whereas nanostructure SnTe is mostly synthesized by ball milling processes [23, 25, 26].

### 2.1 Nano-structure SnTe synthesis

Nano-structure TE materials may be synthesized using two different approaches, one is top-down methods and another is bottom-up techniques. Top-down approaches are referred as physical method whereas bottom-up techniques may be physical, chemical or both methods. The starting materials of top-down approaches may be solid in general. However, bottom-up approaches are started with liquid or gas phase of materials. Possible methods for solid phase synthesis of nano-structured materials are ball milling, Lithography, severe plastic deformation etc. [30, 31] Hydrothermal method, Sol-gel method, co-precipitation method, Sono-chemical method, micro-emulsion method, and biometric method etc. may be used to synthesis nano-structured materials, using liquid phase. The methods by employing gas phases to synthesis materials are chemical vapor deposition, pulsed laser synthesis, spray pyrolysis method, sputtering method, and inert gas condensation etc. [30, 31].

## 3. Power factor optimization

ZT of TE materials plays an important role to determine the efficiency of TE device, where $ZT = \left(\frac{S^2\sigma}{\kappa_e + \kappa_L}\right)T$ ) and PF=$S^2\sigma$. Further, output power density is directly related with PF of the material [2]:

$$\omega_{max} = \frac{1}{2}\frac{(T_H - T_C)^2}{L}S^2\sigma \qquad (1)$$

Where, $T_H$, $T_C$ and L are the hot side temperature, cold side temperature and length of leg of TE device respectively. The maximum output power density is directly related to PF. Hence, high efficiency and large output power may be obtained by tuning PF of a TE material [2]. The equations,

$$S = \frac{8\pi^2}{3}\frac{K_B^2}{eh^2}m^* T \left(\frac{\pi}{3n}\right)^{2/3} \qquad (2)$$

$$\sigma = ne\mu \qquad (3)$$

$$k_e = L\sigma T \tag{4}$$

determine the relation among $S$ and $\sigma$, and n along with density of states effective mass ($m^*$). However, DOS effective mass $m^*$ is strongly correlated with band effective mass $m_b^*$ and valley degeneracy $N_v$ through the relation [2]:

$$m^* = N_v^{2/3} m_b^* \tag{5}$$

The band effective mass $m_b^*$ may be tuned, applying band engineering process viz., band convergence, resonant state, band flattening and band inversion etc. However, PF also may be enhanced through optimization of n along with band engineering process.

### 3.1 Carrier concentration optimization

Sn vacancies in synthesized SnTe samples is the sole responsible for large hole concentration ($10^{20}$ - $10^{21}$ cm$^{-3}$) in the material [6]. The liquidus line shape in SnTe phase diagram indicates reasons for large Sn vacancies in the synthesized samples [32]. Hence, pristine SnTe shows P-type conductivity. The pristine SnTe possesses a high $\sigma$ value of $\approx 7000\ S\ cm^{-1}$ but extremely low $S$ of $20\ \mu V K^{-1}$ at room temperature [33]. However, suppression of high hole concentration in pristine SnTe may enhance the TE performance of the material. Effect of carrier concentration on S and corresponding PF is presented for few doped samples in Figure 1.

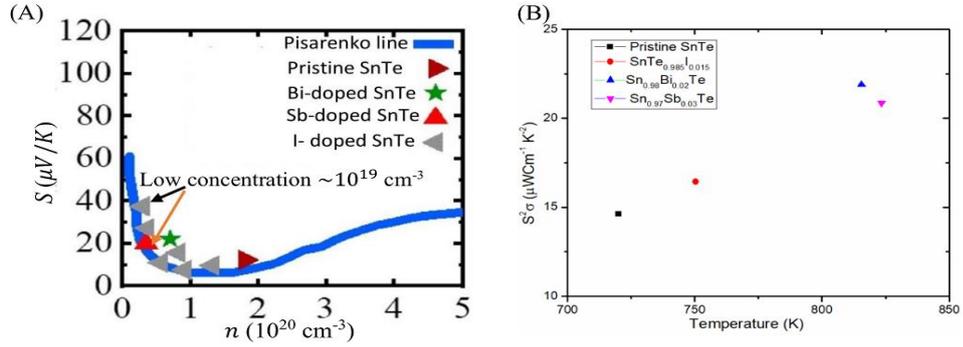

Figure 1. (A). Effect of carrier concentration on Thermopower (*S*), Pisarenko plot. (B) Corresponding, Power factor vs. temperature for un-doped and Bi, Sb I- doped SnTe [3, 33, 34, 35].

### 3.2 Band structure engineering

Band structure engineering is employed to tune various transport parameter of SnTe with the target to enhance PF of the material. The band engineering method may be used to increase equivalent degenerate valleys of the band structure, causes lager S value in a TE material. Band engineering method increases the degenerated valleys through the convergence of different bands in Brillouin zone. In pristine SnTe, contribution of heavy hole in valance band is negligible during the electron-hole transport owing tolarge band gap between L and, Σ bands, leads low *S* value. In this situation band engineering plays an important role, convergence of two valence band increase degeneracy and enhances *S* value of SnTe. Band structure engineering for SnTe includes,

**i. tuning of carrier effective mass:** Another important parameter to tune S as well as PF is carrier effective mass (m*). Eq. (2) indicates the relation between carrier effective mass (m*) and S [23]. Further, Eq. (5) shows that m* is related with number of degenerated valleys ($N_v$) of the band structure and the band effective mass ($m_b$*). Hence, the *S* may be increased by increasing $N_v$ and $m_b^*$ simultaneously. It is noteworthy to mention that mobility ($\mu$) is inversely proportional to $m_b^*$ through following relationship, $\mu = (m_I^*)^{-1}(m_b^*)^{-1}$, where $m_I^*$ represents transport mass. Hence, high $m_b^*$ and lower $\mu$ yield large S. However, low $\mu$ causes less $\sigma$ value, causes reduction in ZT. Optimization between these two parameters is required to enhance PF and ZT [23].

ii. **Resonant level (RL):** the *S* of TE materials may be enhanced by increasing local DOS at Fermi level. The relation between the local DOS with *S* is expressed by Mott expression for degenerate semiconductors [2],

$$S = \frac{\pi^2}{3}\frac{K_B^2}{q} \; T \left[ \frac{1}{n}\frac{dn(E)}{dE} + \frac{1}{\mu}\frac{d\mu(E)}{dE} \right]_{E = E_F} \quad (6)$$

Where, $n(E)$, $\mu(E)$, q, $K_B$, and $E_F$ are the carrier density, carrier mobility, carrier charge, Boltzmann constant and fermi energy, respectively. Hence, Eq. (6) indicates that S may be increased by increasing $n(E)$ at fermi level. This technique may be employed in SnTe through resonant doping, attributes the distortion of DOS at fermi-level.

iii. **Valance band convergence:**

Band degeneracy, effective to enhance TE properties may be achieved by tuning the valence band convergence. Band convergence is realized by merging two or more conduction bands, during chemical doping [36]. It is one of the significant methods to achieve high PF of SnTe based TE material.

iv. **Synergistic effect:**

The band convergence and resonant level are very much effective to enhance power factor. But, two processes are effective at different temperature ranges. However, combination of these two effects in a single compound may improve *S* over the entire working temperature range, is known as synergistic effect.

v. **Band inversion:**

Band inversion method is one of the recent techniques, used for band structure reconstruction in TE materials. Band splitting at some point of Brillouin zone by electron spin-orbit coupling or by lattice strain has been achieved in band inversion method.

## 4. Reduction of lattice thermal conductivity:

Optimization of PF along with reduction of κ is the basic technique to enhance ZT. Hence, TE properties of SnTe may be improved by reducing the electronic ($κ_e$) and lattice ($κ_L$) parts of thermal conductivity, alike other TE materials. $κ_e$ and $κ_L$ are decoupled and may be tuned independently. Optimization of electronic properties in SnTe through RL and band convergence process, and tuning of $n$ optimize the $κ_e$. Hence, $κ_L$ is only TE parameter may only be used to enhance ZT independently, without disturbing other electronic parameters. According to the ideal gas model, the expression of the $κ_L$ may be written as [2],

$$κ_L = \frac{1}{3} C_v v_g^2 \, \tau \quad (7)$$

Here, $C_v$, $v_g$ and $\tau$ are heat capacity, phonon group velocity, and phonon relaxation time respectively. Reduction of $κ_L$ through reduction of $C_v$ is very difficult, because it is an inherent material's property. However, $\tau$ may be minimized by enhancing phonon scattering, either by crystal imperfection or acoustic-optical phonons interaction. It may be achieved through inclusion of point defects, dislocations, grain boundaries, nanostructures and soft phonon modes etc. [2]. And the $v_g$ may be reduced by lattice softening. [2]

### 4.1. All-Scale Hierarchical Architectures:

All-Scale hierarchical architecture is a technique to incorporate or combine point defects and nano-structure during the synthesis of TE material. The prime target is to decrease $\kappa_L$ through phonon scattering at the atomic scale, nanoscale, and mesoscale.

### Summary

SnTe is a lead-free and environmentally friendly alternative potential TE material, which may replace well-known PbTe-based TE materials. Multiple valence bands in SnTe are an advantage to improve ZT, using an additional degree of freedom. Sn vacancies in SnTe exhibit very high hole concentration, leads very poor TE performance. Optimization of $n$ and band engineering by employing an increase in carrier effective mass, RL, band convergence, and band inversion techniques are applied to increase PF and ZT. Optimization of n may be achieved through doping various elements viz., Sb and I in SnTe. Sharp increase in DOS near the Fermi level of SnTe-based TE material, i.e., RL may be achieved by In doping in SnTe. Band convergence is another method to enhance PF. However, synergistic effect, i.e., combination of band convergence and RL are used to enhance ZT at high temperature for SnTe-based TE materials. Point defects and nano-structuring techniques are useful to decrease $\kappa_L$ and improve TE performance of SnTe-based materials. Quantum confinement is helpful to enhance PF and decrease $\kappa_L$ by phonon scattering.

### Acknowledgments

This work is supported by the UGC-DAE-CSR Kalpakkam, India (Ref: CRS/2022-23/04/893 and Ref: CRS/2021-22/04/639) in the form of a sanctioned research project.